\documentclass[11pt,letterpaper]{article}

\usepackage[utf8]{inputenc}
\usepackage{amsmath,amssymb,pdfpages,geometry,nicematrix,graphbox,graphicx,placeins, mathtools, tikz-cd, upgreek, booktabs, multirow, siunitx} 
\usepackage[hidelinks]{hyperref}
\usepackage{appendix}
\usepackage{float}
\usepackage{xfrac}

\usepackage{microtype}

\usepackage[labelfont=bf]{caption}
\usepackage{pgfplotstable}
\pgfplotsset{compat=1.17}

\usepackage[style=science,articletitle=true,doi=false,isbn=false,url=false,eprint=false,maxbibnames=99]{biblatex}
\AtEveryBibitem{\clearfield{month}}
\AtEveryBibitem{\clearfield{day}}

\AddToHook{cmd/cite/before}{\ifhmode\unskip~\fi}

\usepackage{jabbrv}

\renewcommand{\vec}[1]{\mathrm{\mathbf{#1}}}
\newcommand{\angstrom}{\textup{\AA}{}}

\title{Bayesian electron density determination from sparse and noisy single-molecule X-ray scattering images}
\author{Steffen Schultze$^1$, Helmut Grubmüller$^{2*}$}
\date{%
    \textit{Max Planck Institute for Multidisciplinary Sciences}\\[2ex]
    $^1$sschult@mpinat.mpg.de \,\,\,$^2$hgrubmu@mpinat.mpg.de\\[2ex]
    \today
}

\begin{document}

\maketitle

\begin{center}
\small    
\textbf{Short title} \\ Bayesian single-molecule X-ray scattering

\textbf{Teaser} \\ Electron densities can be determined from noisy images without orientation determination
\end{center}

\begin{abstract}
Single molecule X-ray scattering experiments using free electron lasers hold the potential to resolve both single structures and structural ensembles of biomolecules.
However, molecular electron density determination has so far not been achieved due to low photon counts, high noise levels and low hit rates. 
Most analysis approaches therefore focus on large specimen like entire viruses, which scatter substantially more photons per image, such that it becomes possible to determine the molecular orientation for each image. 
In contrast, for small specimen like proteins, the molecular orientation cannot be determined for each image, and must be considered random and unknown. 

Here we developed and tested a rigorous Bayesian approach to overcome these limitations, and also taking into account intensity fluctuations, beam polarization, irregular detector shapes, incoherent scattering and background scattering. 
We demonstrate using synthetic scattering images that it is possible to determine electron densities of small proteins in this extreme high noise Poisson regime. 
Tests on published experimental data from the coliphage PR772 achieved the detector-limited resolution of $9\,\mathrm{nm}$, using only $0.01\,\%$ of the available photons per image. 

\end{abstract}

\clearpage

\section{Introduction}
\begin{figure}
    \centering
    \includegraphics[width=\textwidth]{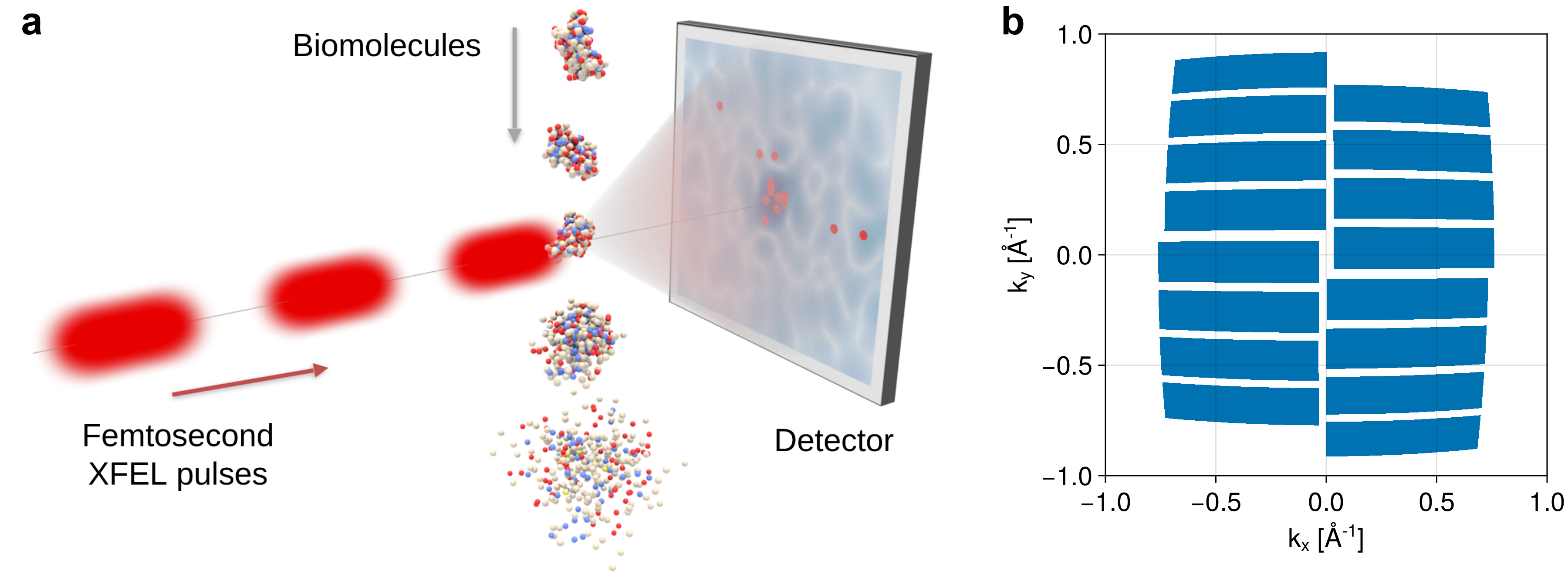}
    \caption{\textbf{a} Single molecule scattering experiment (image reproduced from von Ardenne et al. \cite{von_ardenne_structure_2018}). \textbf{b}~Irregular detector shape used for our simulated scattering experiments, modelled after the detector used at the European XFEL \cite{allahgholi_adaptive_2019}. Note that the apparent `curvature' does not reflect the actual detector geometry, but is instead an artifact of the projection onto the Ewald sphere.}
    \label{fig: introexperiment}
\end{figure}

Ultrashort pulse X-ray scattering experiments using X-ray free electron lasers (XFELs) offer the possibility to take 'snapshots' of biomolecular structures with sub-nanometer spatial and femtoseconds time resolution \cite{hajdu_single-molecule_2000, huldt_diffraction_2003, gaffney_imaging_2007, miao_beyond_2015}. In these `diffraction before destruction' experiments \cite{chapman_diffraction_2014} (Fig.~\ref{fig: introexperiment}a), a stream of sample particles is hit by a series of high intensity, ultra-short (femtoseconds) X-ray pulses; and for each pulse the scattered photons are recorded as a scattering image. Crucially, the pulses are so short that scattering outruns sample 
destruction~\cite{neutze_potential_2000}.

Currently, most of these experiments focus on nano-crystals. These serial femtosecond crystallography experiments have provided both static \cite{schlichting_serial_2015, chapman_structure_2017} as well as time-resolved structures at resolutions better than $3\,$\AA{} \cite{oda_time-resolved_2021,tenboer_time-resolved_2014,barends_serial_2022}. Despite these successes, the need to grow sufficiently well-ordered crystals and the inevitable ensemble averaging pose severe limitations \cite{barends_serial_2022}.

The `holy grail' is therefore to perform X-ray scattering experiments on single non-crystalline particles or even single molecules such as proteins \cite{neutze_potential_2000,gaffney_imaging_2007}. 
Although the high repetition rates of current XFELs of up to $27\,\mathrm{kHz}$ \cite{zastrau_high_2021} allow the collection of millions of such images even at rather low hit rates \cite{sun_current_2018}, substantial challenges remain, such as low photon counts, unknown sample orientation, and low signal to noise ratios. 
These have so far limited such experiments to relatively large particles such as viruses at moderate resolutions of ca.\ $10\,$nm \cite{seibert_single_2011, ekeberg_three-dimensional_2015, hosseinizadeh_high-resolution_2014}. 
Here we address these challenges using a rigorous Bayesian approach, and demonstrate that de novo electron density determination should be possible also for single molecules. 

The first challenge is that, due to the small molecular size, the number of scattered photons is typically very low \cite{neutze_potential_2000}. 
In fact, for single proteins, and despite the high photon flux provided by the XFEL, only ten to several hundred recorded photons per scattering image are expected \cite{von_ardenne_structure_2018}. 
In this extreme Poisson regime, each scattering image thus does not reveal the full scattering intensity distribution (Fig.~\ref{fig: introexperiment}a, light blue color on the detector), but rather consists of only a few discrete photon positions (red dots), distributed according to the unknown scattering intensity. 


The second challenge is that, for each hit, the sample orientation is different, random, and unknown, preventing a naive approach based on averaging many images. 
Several new methods have been developed over the past 20 years to overcome this challenge. 
Most methods aim to determine the molecular orientation from the positions of the scattered photons separately for each image. 
Subsequently, the properly oriented images are assembled in Fourier space into a full three-dimensional scattering intensity \cite{shneerson_crystallography_2008, loh_reconstruction_2009, walczak_bayesian_2014, kassemeyer_optimal_2013, elser_three-dimensional_2011, tegze_atomic_2012, flamant_expansion-maximization-compression_2016, ayyer_dragonfly_2016}, from which the electron density is derived using established phase retrieval methods \cite{elser_searching_2007, luke_relaxed_2004, fienup_phase_1982}.
These orientation determination approaches, including the EMC-algorithm \cite{ayyer_dragonfly_2016, sun_current_2018, hosseinizadeh_conformational_2017} and manifold embedding algorithms such as diffusion map \cite{fung_structure_2009,schwander_symmetries_2012, giannakis_symmetries_2012, winter_enhancing_2016}, typically require $10^2$ to $10^4$ coherently scattered photons per image to determine the sample orientation with sufficient accuracy. 
Moreover, they are rather sensitive to noise, precluding their application to single molecules. 
In fact, for realistic noise levels and photon counts, the information content per scattering image is far too low for successful orientation determination of individual images \cite{walczak_bayesian_2014}.

To circumvent this problem, it has been proposed to extract only orientation-invariant quantities from the scattering images, in particular correlations \cite{saldin_structure_2009, saldin_reconstructing_2011, saldin_beyond_2010, saldin_new_2011, saldin_structure_2010, starodub_single-particle_2012, kurta_correlations_2017, donatelli_iterative_2015, von_ardenne_structure_2018}. 
Notably, using three-photon correlations, it has been shown that density determination should be possible from as few as three recorded photons per image \cite{kam_reconstruction_1980, von_ardenne_structure_2018} as the ultimate limit. 
However, by neglecting higher correlations, much of the scattering information is discarded \cite{von_ardenne_structure_2018}.

The third challenge is posed by several additional sources of experimental noise and uncertainties, mainly due to incoherent scattering, background scattering, and beam intensity fluctuations as well as incomplete and irregular coverage of the scattering solid angle by the detector (Fig.~\ref{fig: introexperiment}b) \cite{von_ardenne_structure_2018, walczak_bayesian_2014, huldt_diffraction_2003, loh_reconstruction_2009}. 
These noise sources are particularly prohibitive at the single particle level, and the usual subtraction of an estimated background noise level and fortuitous error cancellation through averaging fail in this extreme Poisson regime \cite{sun_current_2018}. 
To illustrate this challenge, Fig.~\ref{fig: crambinrealistic} shows a few simulated images at a realistic noise level of $75\%$ to $90\%$ \cite{yoon_comprehensive_2016,sobolev_megahertz_2020,sun_current_2018}.
Due to the lack of a combined, systematic treatment of all three sources of uncertainty, \emph{de novo} electron density determination of single proteins has so far been out of reach.

The Bayesian method we developed and assessed here approaches the problem from a different angle.
Rather than attempting to orient each individual scattering image, the Bayesian posterior probability \emph{given the whole set of images} (typically millions) is considered and either sampled or maximized. 
As an important advantage, this Bayesian approach allows for a systematic inclusion of noise in terms of a physics-based forward model of the scattering experiment and its uncertainties. 
Further, and in contrast to correlation based methods, the full information content of all scattering images is used, thus reducing the required number of images to achieve a particular resolution. 
Finally, posterior sampling provides error bounds and uncertainty estimates for the obtained electron density. 

We tested our approach both on noisy synthetic scattering images as well as on downsampled experimental images. Even for a small single protein (crambin), resolutions of $8\,\angstrom$ to $10.4\,\angstrom$ were achieved under realistic conditions, and up to $4.2\,\angstrom$ under noise-free conditions. 
As a test using experimental images, we successfully recovered the electron density of the coliphage PR772 \cite{reddy_coherent_2017} at $9\,\mathrm{nm}$ detector limited resolution using only $0.01\,\%$ of the recorded photons per image.

\clearpage

\section{Results}
\paragraph{Bayesian inference of single-molecule X-ray scattering.}
We first summarize the Bayesian formalism and our approach. 
For each scattering event $j=1\ldots N$, the positions of the $n_j$ scattered photons are recorded on the detector as a scattering image and are converted into scattering vectors $\vec k_1^{(j)}, \dotsc, \vec k_{n_j}^{(j)}$. 
For each possible electron density function $\rho$, a Bayesian posterior probability is calculated given the set of scattering images $\mathcal I=\{\vec k_1^{(j)}, \dotsc, \vec k_{n_j}^{(j)}\}_{j=1\ldots N}$,
\begin{equation}
    P(\rho\,|\,\mathcal I) \propto P(\mathcal I \,|\, \rho) P(\rho)\,,
\end{equation}
from which the most probable electron density as well as its uncertainty is derived.

The likelihood function $P(\mathcal I \,|\, \rho)$ contains an appropriate physical forward model of the scattering process, including noise, intensity fluctuations, polarization, and irregular detector shapes (see Methods Section). 
Because each image is an independent event, the likelihood decomposes into a product of the likelihoods of each single image $j$,
\begin{equation}\label{eq: likelihood product}
    P(\mathcal I \,|\, \rho) = \prod_{j=1}^N P(\vec k_1^{(j)}, \dots, \vec k_{n_j}^{(j)} \,|\, \rho).
\end{equation}
In the absence of information on the orientation of the sample molecule during each scattering event, the single-image likelihood function is given by marginalizing, i.e., as an average over all possible orientations $\vec R$ of the corresponding conditionalized probability of the scattering image $j$,
\begin{equation}
    P(\vec k_1^{(j)}, \dots, \vec k_{n_j}^{(j)} \,|\, \rho) 
    = \int_{\mathrm{SO}(3)}\!\!\! P(\vec k_1^{(j)}, \dots, \vec k_{n_j}^{(j)} \,|\, \rho,{\vec R})\,\mathrm{d}\vec R\,.
\end{equation}
Here, $\mathrm{SO}(3)$ denotes the set of all three-dimensional rotation matrices. 
The latter was calculated from the forward scattering model as described in the Methods Section. To maximize or sample from the Bayesian posterior, a combined Markov chain Monte Carlo (MCMC) simulated annealing approach was used. 

As a physics-motivated real space representation of the electron density $\rho$, we chose a sum of Gaussian functions, each of which may represent an atom, an amino acid, or a larger domain of the sample, depending on the target resolution. 
This choice serves both to minimize the number of required degrees of freedom and as a means of regularization.
Notably, by using such a representation for the electron density as opposed to its Fourier transform, our approach circumvents the phasing problem. 

For a typical target protein consisting of up to several hundred amino acids, the number of required degrees of freedom remains nevertheless large and poses a formidable sampling challenge.
To achieve sufficient sampling, we applied a hierarchical simulated annealing approach as described in the Methods section. 
Briefly, starting at very low resolution and correspondingly few Gaussian functions, the electron densities were sampled in multiple hierarchical stages of increasing resolution; and in each of these stages, the previous electron density of maximal posterior probability was used as a proposal density for the MCMC steps.
\filbreak 

\begin{figure}[h!]
    \centering
    \includegraphics[width = \textwidth]{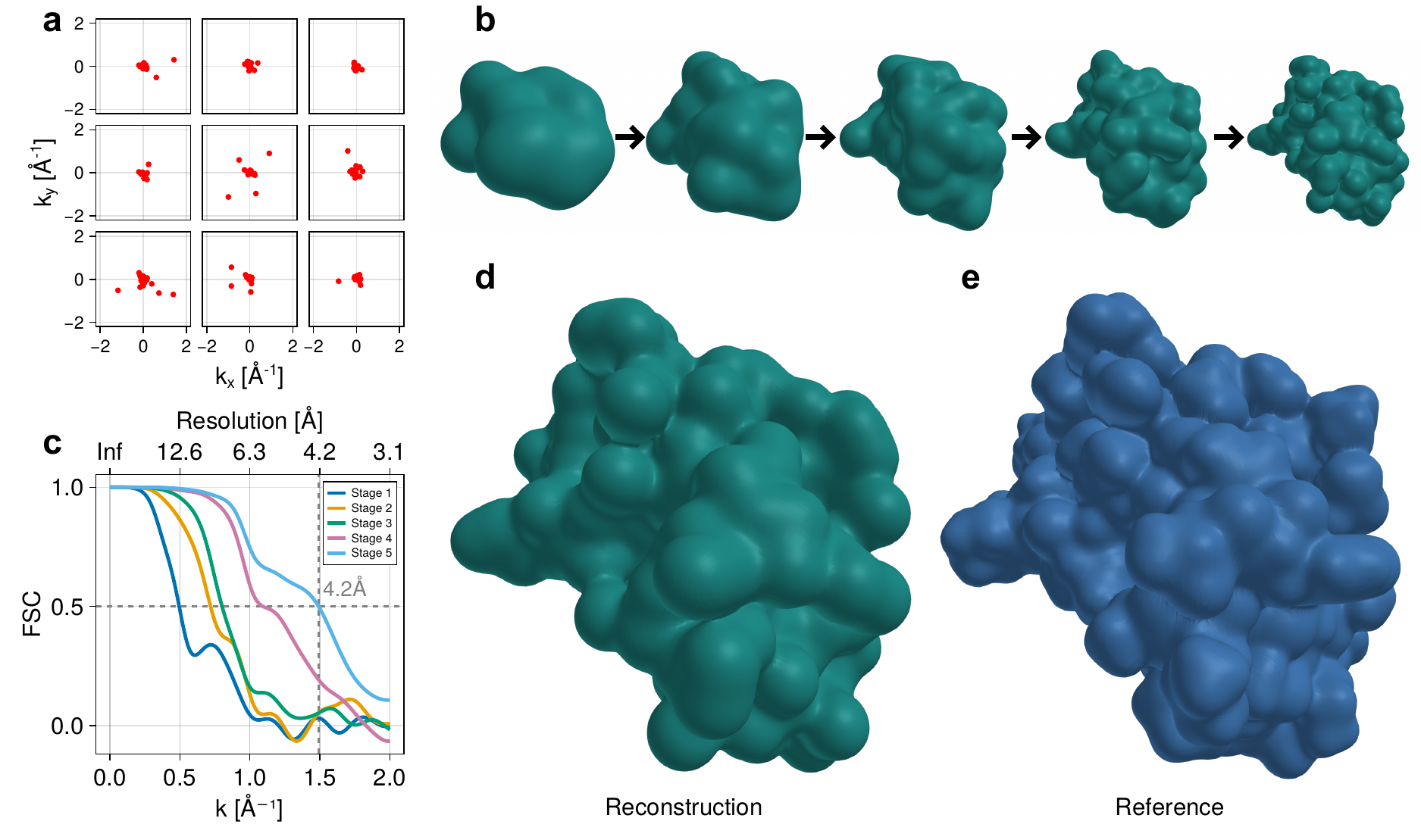}
    \caption{Electron density determination from noise-free images. \textbf{a} Sample synthetic noise-free images, containing only coherent signal photons (red dots). \textbf{b} Hierarchical stages of retrieved electron densities. \textbf{c}~Fourier shell correlation between reconstructed and reference density. \textbf{d}~Reconstructed electron density. \textbf{e} Reference electron density.}
    \label{fig: crambin}
\end{figure}

\paragraph{Electron density reconstruction from noise-free images.}
We first tested our method on synthetic noise-free images, using the same 46 amino acid protein crambin \cite{jelsch_accurate_2000} that was used for the assessment of previous correlation based methods \cite{von_ardenne_structure_2018}. 
Because our Bayesian approach uses all available information, we expect it to require fewer scattering images to achieve the same resolution. 
To test this expectation, a total of $10^8$ noise-free synthetic scattering images were generated, containing a realistic average number of $15$ photons each \cite{von_ardenne_structure_2018,hantke_condor_2016}. 
As described in the Methods section, for each image a random molecular orientation was chosen, and for each orientation, the number of photons per image was drawn from a Poisson distribution. 
Here an intensity of $10^{12}$ photons per pulse was assumed with a beam diameter of $1\,\mathrm{\text{\textmu} m}$ \cite{yoon_comprehensive_2016}. 
Figure~\ref{fig: crambin}a shows several of these images as an example.
\filbreak 

From these images, the electron density was determined in five hierarchical stages (Fig.~\ref{fig: crambin}b), increasing the number of Gaussian functions representing the electron density $\rho$ by a factor of two in each stage. 
For the final stage, $184$ Gaussian functions were used, which is four times the number of amino acids. For more details, see Supplementary Table S1. 
Indeed, a similar Fourier shell correlation resolution \cite{van_heel_fourier_2005} of $4.2\,\angstrom$ (Fig.~\ref{fig: crambin}c) was obtained as in the previous study of our group~\cite{von_ardenne_structure_2018} using only half the number of scattered photons.
Here the Fourier shell correlations serve to compare the reconstructed electron density map (Fig.~\ref{fig: crambin}d) with the reference one (Fig.~\ref{fig: crambin}e) that was used to generate the synthetic scattering images. 
As a further measure of quality, the optimal transport plan between the reconstructed and reference electron densities was computed using a standard algorithm \cite{cuturi_sinkhorn_2013}, obtaining an earth mover's distance of $1.45\,${\AA}.

\filbreak

\begin{figure}[h]
	\centering
	\includegraphics[width=\textwidth]{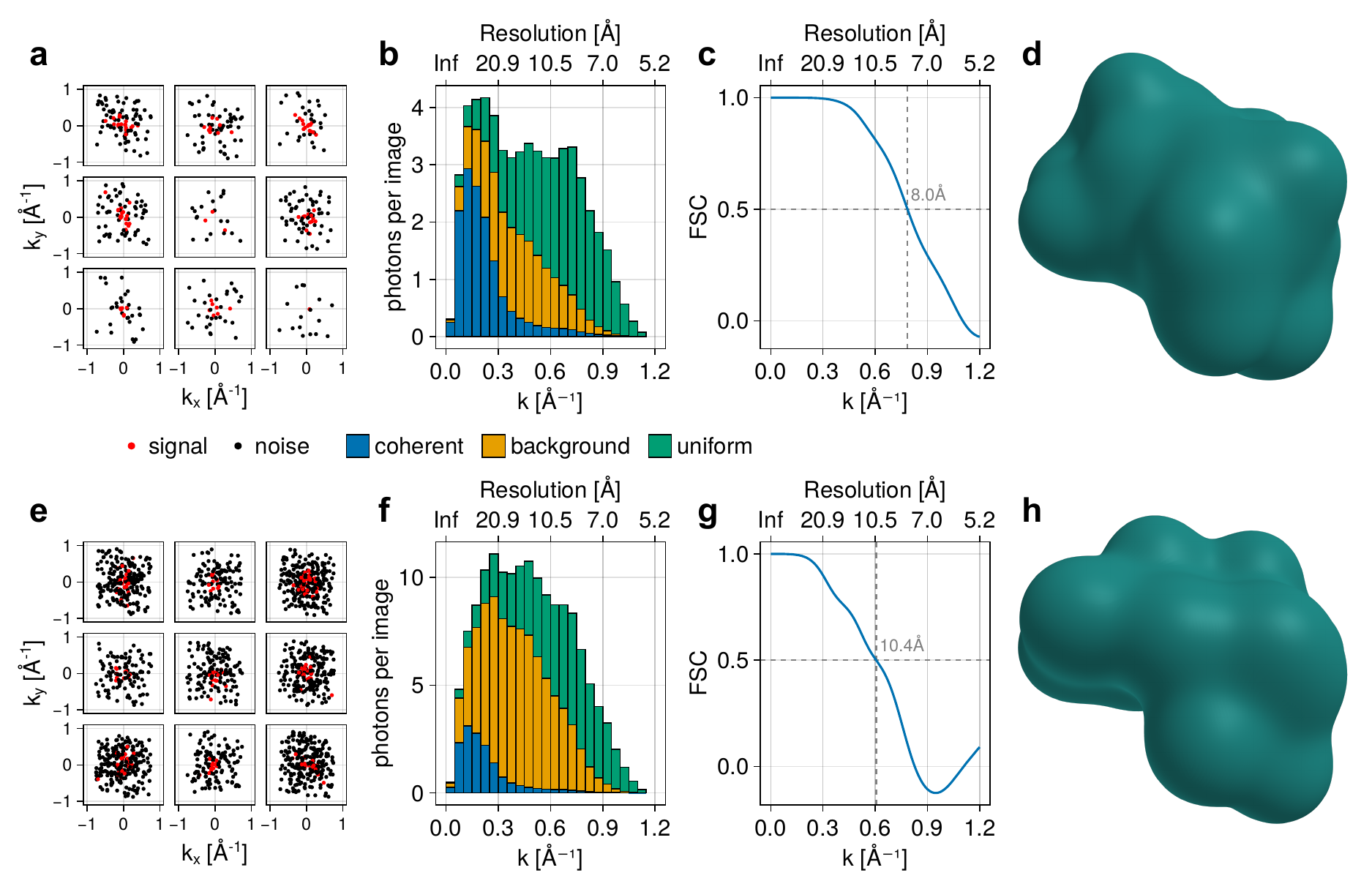}
	\caption{Electron density determination from noisy images, for \textbf{a}-\textbf{d} a noise level of $60\,\%$ and \textbf{e}-\textbf{h} of $90\%$. \textbf{a,e} Sample synthetic noisy images, showing coherent signal photons (red) and noise photons (black). \textbf{b,f} Radial distribution of photons from coherent scattering, background and incoherent (uniform) noise (stacked histogram). \textbf{c,g} Fourier shell correlations show the achieved resolutions of $8\,\angstrom$ and $10.4\,\angstrom$. \textbf{d,h} Reconstructed electron densities.}
    \label{fig: crambinrealistic}
\end{figure}

\paragraph{Density determination from noisy images.}
To assess our method also in the presence of realistic noise sources, we tested it on synthetic noisy scattering images for the same protein Crambin \cite{jelsch_accurate_2000}. 
Because estimates for the experimentally achievable noise level vary and depend on the exact experimental setup \cite{yoon_comprehensive_2016, sobolev_megahertz_2020, sun_current_2018}, two different noise levels were considered:
One million synthetic images were generated at a noise level of $75\,\%$ (Fig.~\ref{fig: crambinrealistic}a-d), and three million at a noise level of $90\,\%$ (Fig.~\ref{fig: crambinrealistic}e-h),
which both are within experimental reach \cite{yoon_comprehensive_2016, sobolev_megahertz_2020, sun_current_2018}. 
Figures~\ref{fig: crambinrealistic}a and \ref{fig: crambinrealistic}e show representative examples of these images. 
Here, the same average number of $15$ coherently scattered signal photons (red) per image as above was assumed. 
Only these contain structural information, but are indistinguishable from the noise photons (black). 
In addition, a total of on average $44$ (Fig.~\ref{fig: crambinrealistic}a) or $137$ (Fig.~\ref{fig: crambinrealistic}e) photons per images of all detected photons were assumed to be incoherent. 
As shown in Fig.~\ref{fig: crambinrealistic}b and \ref{fig: crambinrealistic}f, these were assumed to be from background scattering on carrier gas or solvent molecules (orange, described here by a normal distribution with standard deviation $0.35\,\angstrom^{-1}$), as well as from incoherent scattering (green, described by a uniform distribution). For more details see Methods section and Supplemental Table 1. 

Despite these low signal-to-noise ratios, Fig.~\ref{fig: crambinrealistic}c and \ref{fig: crambinrealistic}g show that structural information is recovered at a conservative Fourier shell correlation resolution estimate of $8\,\angstrom$ in the case of $75\,\%$ noise and $10.4\,\angstrom$ in the case of $90\,\%$ noise.
Figure~\ref{fig: crambinrealistic}d and \ref{fig: crambinrealistic}f show the obtained electron densities, that represent the overall shape of the molecule at the two resolutions. 

\begin{figure}
    \centering
    \includegraphics[width = \textwidth]{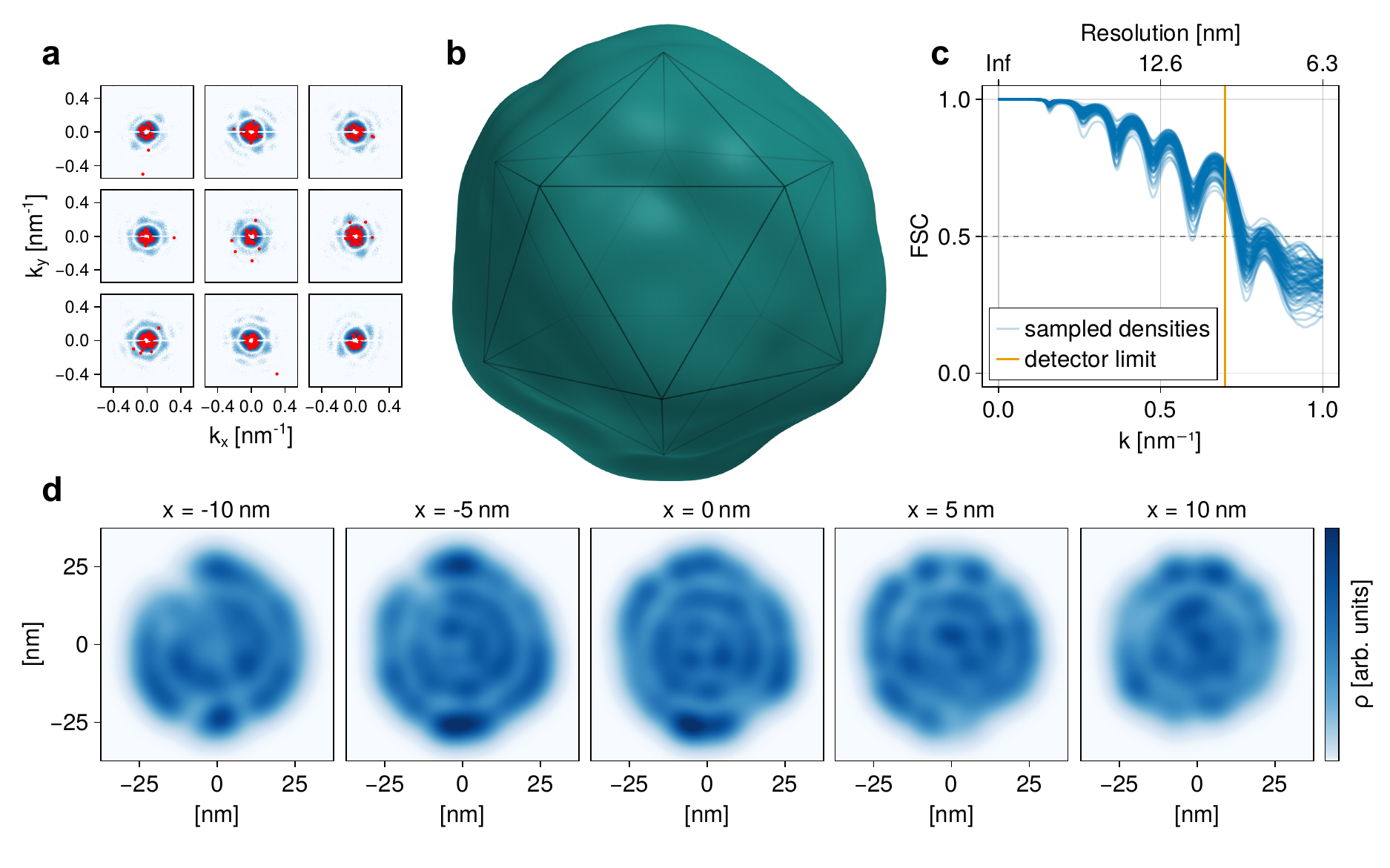}
    \caption{Electron density determination of the coliphage PR772. \textbf{a} Sample downsampled images (red dots) and corresponding original experimental images (blue, log-scaled). \textbf{b} Average density obtained from MCMC-sampling. Perfect icosahedron with side length $30\,\mathrm{nm}$ for reference. \textbf{c} Fourier shell correlations between $100$ randomly selected sampled electron densities and the average density. \textbf{d} Slices through the average electron density on the $yz$-plane for various values of $x$.}
    \label{fig: coliphage}
\end{figure}

\paragraph{Application to experimental data.}
Having assessed our method using synthetic data for which the `ground truth' is known, we next tested it on published experimental data for the icosahedral coliphage PR772 \cite{reddy_coherent_2017}. 
Because this virus is much larger than the protein molecule considered above, on average about $400\,000$ photons per scattering image were recorded. 
Notably, only photons up to $k=0.69\,\mathrm{nm}^{-1}$ were recorded, which limits the resolution to $9\,\mathrm{nm}$. 
For a fair assessment, and to mimic the more challenging low photon counts expected for single-molecule scattering experiments, we downsampled the original images by a factor of $10^4$ using rejection sampling to obtain images with an average of $40$ photons per image (Fig.~\ref{fig: coliphage}a). 
A data set consisting of $2\cdot 10^5$ such images was emulated by randomly downsampling the same images multiple times. 
Because the intensity distribution of the incoming beam is unknown due to the applied hit selection, intensity fluctuations were taken into account via normalization as described in the Methods section.

The virus electron density was described by $n=200$ Gaussians functions of width $\sigma=3.5\,\mathrm{nm}$, adapted to the resolution set by the experimental data. 
To better represent the virus electron density at this scale, both the positions $\vec y_i$ and the heights $h_i$ of the Gaussian functions were considered unknown and determined during the annealing. 
After the annealing procedure, the density was further sampled for $2\cdot 10^6$ MCMC steps.
The thus obtained sample of electron densities was rotationally aligned and averaged. 
The resulting density (Fig.~\ref{fig: coliphage}b) resembles the expected icosahedral structure of the virus. 
As shown in Fig. \ref{fig: coliphage}d, even the internal structure consisting of multiple concentric shells was resolved.
An animated version of Fig.~\ref{fig: coliphage}b is provided as Supplementary Movie S1. 

To obtain an error estimate, the Fourier shell correlations of a randomly selected small subset of these sampled electron densities relative to the averaged density were computed (Fig.~\ref{fig: coliphage}c). 
As can be seen, the $9\,\mathrm{nm}$ resolution limit imposed by the detector geometry was largely achieved. 
In contrast to Hosseinizadeh et al. \cite{hosseinizadeh_conformational_2017}, icosahedral symmetry was not imposed, and therefore the electron density determined here deviates from a perfect icosahedral symmetry (as is also obvious from the slices shown in Fig.~\ref{fig: coliphage}d), which has also been observed for this data set using other methods \cite{ayyer_low-signal_2019,kurta_correlations_2017}.


\FloatBarrier

\section{Discussion}
We have demonstrated electron density determination from highly noisy and sparse single particle X-ray scattering images using a rigorous Bayesian approach. 
Uncertainties such as unknown sample orientation, beam intensity fluctuations, polarization, irregular detector shapes, Poisson noise due to the typically very few recorded photons per image, as well as noise from incoherent and background scattering have been taken into account using a realistic physics-based forward model. 
This model can be adapted to specific experimental conditions and can be generalized to include other noise sources such as detector noise as well. 

Our simulated scattering experiments demonstrate that electron densities can be reliably determined even in this high noise regime well beyond $1\,\mathrm{nm}$ resolution. 
There is no fundamental limit to achieving even higher resolutions, given sufficient numbers of scattering images. In contrast, for approaches based on orientation determination, the resolution is limited by the number of photons per image \cite{walczak_bayesian_2014}. 


Because our approach uses all available information, fewer scattering images and fewer photons per image are required to achieve the same resolution than by previous approaches, in particular compared to correlation methods \cite{von_ardenne_structure_2018}. 
Remarkably, for the coliphage test case, $10^4$ times downsampled images sufficed to recover the icosahedral structure at the detector-limited resolution of $9\,\mathrm{nm}$.
In fact, $10$ to $100$ times fewer photons per image were required than what was so far considered the `low signal limit' \cite{ayyer_low-signal_2019}. 
For this comparison, note that our photon counts refer to entire images, whereas Ayyer~et~al. \cite{ayyer_low-signal_2019} report photon counts excluding the central speckle that contains most of the photons. 

There is no fundamental resolution limit to our approach. 
However, Bayesian sampling in high-dimensional search spaces generally poses computational challenges. The problem-adapted hierarchical sampling method presented here alleviated this technical limitation markedly and allowed the optimization of electron density representations with up to $600$ degrees of freedom. 
Nevertheless, for increasing sample size to resolution ratios, the main bottleneck of our approach is the computational effort, both due to the required sampling and the large number of scattering images.
Although a brute-force approach is possible due to inherently parallel computations, improved optimization and sampling methods will be helpful to address this issue \cite{luke_stochastic_2024}, as well as the use of prior structural information from, for example, structure databases, AlphaFold \cite{jumper_highly_2021}, or molecular dynamics force fields. 

This computational bottleneck does not preclude the application to larger specimen, like the coliphage studied here. 
Rather, the computational effort increases with the complexity of the used electron density representation. 
Whereas we have, under ideal conditions, demonstrated that a resolution of $4.2\,$\AA{} can be achieved for a small soluble protein, for larger complexes such as the ribosome, substantial computational resources would be required to achieve resolutions better than about $3\,\mathrm{nm}$.

Whereas our results demonstrate that our Bayesian approach should enable structure determination from noisy single-molecule X-ray scattering images, we have so far only assessed its performance and accuracy on synthetic scattering images or on preprocessed images from diffraction experiments on much larger virus specimen.
Because the approach only rests on a physical forward model of the experiment, further sources of noise of experimental uncertainties can be readily implemented in a systematic way. Although the forward model presented in this work turned out to be sufficiently accurate to enable successful reconstruction of the PR772 virus, future calibration and improvements will be beneficial, such as more complex detector models, the effect of a possible solvation shell around the molecule, structural heterogeneity, or the identification of hits vs.\ misses.

\section{Materials and Methods}

\paragraph{Noise-free forward model.}
\label{sec: forward model}
In the experiments, single sample molecules enter a pulsed femtosecond XFEL-beam, and for each pulse, the positions of the scattered photons are recorded on the detector as a scattering image. 
Each location on the detector corresponds to a specific scattering vector $\vec k = \vec k_\mathrm{i} - \vec k_\mathrm{s}$ on the Ewald sphere $E$ in Fourier space, where $\vec k_\mathrm{i}$ is the incident wave vector and $\vec k_\mathrm{s}$ the wave vector after scattering. 
Each scattering image is, therefore, given by a list of scattering vectors $\vec k_1, \dotsc, \vec k_l$. 
Their probability distribution is given by three-dimensional intensity function $I_\rho(\vec R\vec k) = \lvert \mathcal F\{\rho\}(\vec R\vec k)\rvert^2$, which for coherent scattering is given by the Fourier transform of the electron density $\rho$. 
Here, $\vec R \in \mathrm{SO}(3)$ is a rotation matrix describing the orientation of the molecule.

The likelihood that an image $\vec k_1, \dotsc, \vec k_l$ is observed for a given electron density $\rho$ is obtained by averaging the conditional likelihood over all possible orientations $\vec R$. 
This conditional likelihood is given by the product of a Poisson distribution for the number of photons $l$ and, because the photons are conditionally independent given $\vec R$, a product of the intensity function evaluated at the scattering vectors of the scattered photons,
\begin{equation}
    \label{eq: likelihood noise-free}
    \begin{aligned}
        P(\vec k_1, \dots, \vec k_l \,|\, \rho)
        &\propto \int_{\mathrm{SO}(3)}\!\mathrm{d}\vec R\,\, I_0^l \exp\left(-I_0\!\int_\mathrm{E} I_\rho(\vec R\vec k) \,\mathrm d \vec k\right) \prod_{i=1}^l I_\rho(\vec R\vec k_i).
    \end{aligned}
\end{equation}
Note that here and subsequently normalization factors and constants such as the electron radius are omitted; instead the normalization is chosen at the end such that the correct photon counts are obtained. 

This likelihood, given by equations \eqref{eq: likelihood noise-free} and \eqref{eq: likelihood product}, represents the complete noise-free forward model, which forms the basis for the subsequent inclusion of error models. 

\paragraph{Incoherent and background scattering.}
In addition to the coherent photons, also incoherently scattered photons from, for example, Compton scattering and Auger decay, are observed. 
They represent up to $90\%$ of the total scattered photons, but are distributed uniformly on the Ewald sphere. 
They therefore spread over a much larger solid angle than the coherent photons, such that the effective amount of noise due to this incoherent scattering is smaller. 
For this reason, the noise due to incoherent scattering is larger for increasing resolutions, whereas at lower resolutions of about $10\,\mathrm{nm}$ that have been demonstrated for viruses it can be neglected.

A second source of noise is scattering from other molecules, such as water molecules attached to the sample in aerosol delivery \cite{kirian_simple_2015}, bulk water for liquid beam \cite{deponte_gas_2008} or sheet \cite{hoffman_microfluidic_2022} delivery, or remaining gas molecules in the beam volume.
These molecules scatter both coherently and incoherently, but, due to the random positions and orientations of these particles, incoherent summation to $I_\rho$ is a good approximation. 

Neglecting beam polarization for a moment, the distribution of the photons from incoherent and background scattering is radially symmetric.
For simplicity, here a uniform distribution on the Ewald sphere is assumed for the incoherently scattered photons and a Gaussian distribution centered at the origin of reciprocal space for the background scattering.
Other radial distributions, for example from measurements, can of course be readily implemented.

To include incoherent and background scattering within the likelihood function, their distributions are added to the intensity function, replacing $I_\rho$ by
\begin{equation}
    I_\mathrm{n}(\vec k, \rho) = I_\rho(\vec k) + I_\mathrm{b}(\vec k) + I_\mathrm{u}
\end{equation}
in equation \eqref{eq: likelihood noise-free}.
Here, the constant $I_\mathrm{u}$ describes the contribution from uniform incoherent scattering, and
\begin{equation}
    I_\mathrm{b}(\vec k) = \frac{C_\mathrm{b}}{2\pi\sigma^2}\exp\left(-\frac{\vec k^2}{2\sigma^2}\right)
\end{equation}
is the Gaussian distribution describing  background scattering.

\paragraph{Polarization.}
To additionally include the linear polarization of the XFEL beam, the scattering intensity is changed by a factor $f_\mathrm{p}(\vec k) = \cos^2\theta + \cos^2\phi\sin^2\theta = 1 - k_y^2\lambda / 2\pi$, where $\theta$ is the scattering angle and $\phi$ the azimuthal angle relative to the direction of polarization \cite{giacovazzo_fundamentals_2011}. 
As a consequence, for each scattering vector $\vec k$, the expected number of photons from coherent and Compton scattering is reduced by $f_\mathrm{p}(\vec k)$. 
In contrast, the distribution of photons arising from Auger decay is unaffected. 
For our forward model, we assume therefore that the Gaussian noise $I_\mathrm{b}$ is multiplied by this factor while the uniform noise is not. 

Accordingly, $I_\mathrm{n}$ is replaced by
\begin{equation} \label{eq: intensity function}
    I_\mathrm{np}(\vec R, \vec k, \rho) = f_\mathrm{p}(\vec k) (I_\rho(\vec R\vec k) + I_\mathrm{b}(\vec k)) + I_\mathrm{u},
\end{equation}
which now also depends on the molecular orientation. 
As result, rotating the molecule around the beam axis does no longer correspond directly to rotating the scattering images because the polarization orientation is stationary.
The likelihood becomes
\begin{equation}
    P(\vec k_1, \dots, \vec k_l \,|\, \vec \rho) 
    \propto \int_{\mathrm{SO}(3)}\!\mathrm{d}\vec R\,\, I_0^l \exp\left(-I_0\!\int_E \! \mathrm d \vec k \,\, I_\mathrm{np}(\vec R, \vec k, \rho)\right)
    \prod_{i=1}^l I_\mathrm{np}(\vec R, \vec k_i, \rho).
\end{equation}

\paragraph{Irregular detector shape.}
Most X-ray detectors have irregular shapes. For example, the detector used at the European XFEL \cite{allahgholi_adaptive_2019} is composed of $16$ separate modules arranged as shown in F~\ref{fig: introexperiment}b. 
In the forward model, the shape of the detector is encoded in the detection probability $p_\mathrm{d}(\vec k)$ that a photon with scattering vector $\vec k$ is registered by the detector. 
This formalism allows for the inclusion of any detector shape and can also be used to include individual detection probabilities per pixel.

The resulting likelihood function is a straightforward extension similar to the above polarization, the only difference being that here, all photons are affected. 
As a consequence, the factors $p_\mathrm{d}(\vec k_i)$ in the product over $i$ factor out and can be omitted because they do not depend on the images,
\begin{equation}
    \begin{split}
        P(\vec k_1, \dots, \vec k_l \,|\, \vec \rho) 
        &\propto \int_{\mathrm{SO}(3)}\!\mathrm{d}\vec R\,\, I_0^l \exp\left(-I_0\!\int_E \! \mathrm d \vec k \,\, p_\mathrm{d}(\vec k) I_\mathrm{np}(\vec R, \vec k, \rho)\right)
        \prod_{i=1}^l p_\mathrm{d}(\vec k_i) I_\mathrm{np}(\vec R, \vec k_i, \rho) \\
        &\propto \int_{\mathrm{SO}(3)}\!\mathrm{d}\vec R\,\, I_0^l \exp\left(-I_0\!\int_E \! \mathrm d \vec k \,\, p_\mathrm{d}(\vec k) I_\mathrm{np}(\vec R, \vec k, \rho)\right)
        \prod_{i=1}^l I_\mathrm{np}(\vec R, \vec k_i, \rho).
    \end{split}
\end{equation}

\paragraph{Intensity fluctuations.}
Fluctuations of the incoming beam intensity $I_0$ are described by a Gamma distribution $I_0\sim \langle I_0 \rangle \Gamma(\alpha, \beta)$, where $\langle I_0 \rangle$ is the average intensity \cite{yun_coherence_2019,saldin_statistical_1998,ayvazyan_first_2006}. 
The shape and rate parameters $\alpha$ and $\beta$ depend on the specific free electron laser. 
For the forward model, $\alpha = \beta = 4$ was assumed, which as has been determined for an XFEL operating at $32\,\mathrm{nm}$ wavelength \cite{ayvazyan_first_2006}.

To include these fluctuations within the likelihood function, an additional integral over $I_0$ weighted by the probability density of the Gamma distribution is required,
\begin{equation} \label{eq: integral over I_0}
    \begin{split}
        & P(\vec k_1, \dots, \vec k_l \,|\, \vec \rho) \\
        &\quad \propto \int_0^\infty \!\!\mathrm d I_0 \,  I_0^{\alpha-1} \exp\left(-\frac{\beta I_0}{\langle I_0\rangle} \right) \int_{\mathrm{SO}(3)}\!\mathrm{d}\vec R\,\,  I_0^l \exp\left(-I_0\!\int_E \! \mathrm d \vec k \,\, p_\mathrm{d}(\vec k) I_\mathrm{np}(\vec R, \vec k, \rho)\right)
        \prod_{i=1}^l I_\mathrm{np}(\vec R, \vec k_i, \rho).
    \end{split}
\end{equation}
Conveniently, this integral can be carried out analytically, and the likelihood reads
\begin{equation} \label{eq: likelihood full}
    P(\vec k_1, \dots, \vec k_l \,|\, \vec \rho)
    \propto \int_{\mathrm{SO}(3)}\!\mathrm{d}\vec R\,\, \Gamma(l + \alpha) \left(\frac{\beta}{\langle I_0\rangle} + \!\int_E\!\mathrm d \vec k\, p_\mathrm{d}(\vec k) I_\mathrm{np}(\vec R, \vec k, \rho)\right)^{\!\!-l-\alpha} 
    \prod_{i=1}^l I_\mathrm{np}(\vec R, \vec k_i, \rho).
\end{equation}

For an unknown intensity distribution, for example when it is biased by hit selection, intensity fluctuations can be treated by considering the probabilities conditionalized on the number of photons $l$, replacing $P(\vec k_1, \dots, \vec k_l \,|\, \vec \rho)$ with 
\begin{equation} \label{eq: likelihood normalized}
    P(\vec k_1, \dots, \vec k_l \,|\, \vec \rho, l)
    \propto \int_{\mathrm{SO}(3)}\!\mathrm{d}\vec R\,\, \left(\int_E\!\mathrm d \vec k\, p_\mathrm{d}(\vec k) I_\mathrm{np}(\vec R, \vec k, \rho)\right)^{\!\!-l} 
    \prod_{i=1}^l I_\mathrm{np}(\vec R, \vec k_i, \rho)
\end{equation}
in equation \eqref{eq: likelihood product}. As this form is independent of $I_0$, the intensity does not have to be integrated as a nuisance parameter.

The likelihood function in equation \eqref{eq: likelihood full} and \eqref{eq: likelihood normalized} represents the so far complete forward model, including incoherently scattered photons, background scattering, beam polarization, the irregular detector shape and intensity fluctuations. 
Equation \eqref{eq: likelihood full} was used for the tests on synthetic data, and equation \eqref{eq: likelihood normalized} for the analysis of the coliphage PR772 data set. 

\paragraph{Structure representation.}
Electron density functions of the reference structures were described by a sum of $m$ Gaussian beads with positions $\vec y_i$, heights $h_i$ and standard deviations $\sigma_i$,
\begin{equation}\label{eq: representation}
    \rho(\vec r) = \sum_{i=1}^m \frac{h_i}{\left(\sigma_i\sqrt{2\pi}\right)^3} \exp \left(-\frac1{2\sigma_i^2}\lVert \vec r - \vec y_i\rVert^2\right)\,.
\end{equation}
Electron density functions of the determined electron densities were described similarly, with one common standard deviation $\sigma=\sigma_i$ and one common height $h = h_i$ for the tests on Crambin, but independent heights $h_i$ for the test on the coliphage. These were treated as unknowns and determined together with the positions $\vec y_i$.

\paragraph{Simulated scattering experiments.}
The forward model described in Section~\ref{sec: forward model} was simulated by the following procedure. To generate one scattering image from an electron density $\rho$, 
\begin{enumerate}
    \item draw the intensities $I_0 \sim \langle I_0 \rangle\Gamma(\alpha, \beta)$,
    \item draw the orientation $\vec R \sim \mathcal U(\mathrm{SO}(3))$ from a uniform distribution on the rotation group,
    \item draw $\bar l \sim \mathrm{Pois}\left(I_0 4\pi (2\pi/\lambda)^2 I(\vec R, \vec 0, \rho)\right)$, with the intensity function $I = I_\rho$ for noise-free images or $I = I_\mathrm{np}$ from equation \eqref{eq: intensity function} for noisy images,
    \item draw photon positions $\vec k_1, \dotsc, \vec k_{\bar l}$ uniformly distributed on the Ewald sphere, and accept each with probability $p_\mathrm{d}(\vec k_i) I_\mathrm{np}(\vec R, \vec k_i, \rho) / I_\mathrm{np}(\vec R, \vec 0, \rho)$.
\end{enumerate}
Note that this procedure works correctly because the intensity function $I(\vec R, \vec k, \rho)$ is always maximal at $\vec k = \vec 0$. To generate the noise-free scattering images, $I_0$ was instead set as a constant $I_0 = \langle I_0\rangle$, and the detector geometry was given by $p_\mathrm{d}(\vec k) = 1$. 

\paragraph{Computation of likelihoods}
\label{sec: distribution}
The integral over the orientation $\vec R$ in equations \eqref{eq: likelihood noise-free}, \eqref{eq: likelihood full} and \eqref{eq: likelihood normalized} was approximated by averaging over a discrete set of typically $n\approx 10^3$ to $n\approx 10^5$ rotations $\vec R_i$ with weights $w_i$. For example, in the noise-free case equation \eqref{eq: likelihood noise-free} was approximated as
\begin{equation}
    P(\vec k_1, \dots, \vec k_l \,|\, \rho) \approx \sum_{i=1}^n w_i I_0^l \exp\left(-N\!\int_E I_\rho(\vec R_i\vec k) \,\mathrm d \vec k\right) \prod_{j=1}^l I_\rho(\vec R_i\vec k_j).
\end{equation}
The rotations $\vec R_i$ and their weights $s_i$ were constructed by combining a Lebedev quadrature rule on $S^2$ \cite{noauthor_sphere_lebedev_rule_nodate} with a uniform quadrature rule on $S^1$, as described by Gräf and Potts \cite{graf_sampling_2009}. To further increase the computational efficiency, the photon positions $\vec k_i$ were discretized. For a more detailed explanation and further implementation details, see the Supplementary Text.

\paragraph{Monte Carlo simulated annealing.}
Optimization and posterior sampling was performed via Markov chain Monte Carlo (MCMC) simulated annealing \cite{kirkpatrick_optimization_1983}. An exponential temperature schedule $T(t) = T_0\exp(-t \ln 2 \,/\, t_{1/2} )$ was used, with $T_0$ and $t_{1/2}$ as listed in Supplementary Table S1. 
The sampling challenge due to the high number of degrees of freedom at high resolutions was alleviated by determining the electron density in multiple hierarchical stages with an increasing number of Gaussian beads. 
In each stage, the density from the previous stage was used as a proposal density, markedly increasing the sampling performance. 
To define this proposal density, let $\vec y_1, \dots, \vec y_n$ be the positions of the Gaussian functions from the previous stage, and $\vec z_1, \dots, \vec z_{2n}$ those of the current stage. 
Then the proposal density was, up to normalization, given by
\begin{equation}\label{eq: proposal density}
    g(\vec z_1', \dots, \vec z_{2n}' \,|\, \vec z_1, \dots, \vec z_{2n}) \propto \prod_{i=1}^{2n} \exp\!\left(-\frac{\lVert\vec z_i' - \vec z_i\rVert^2}{2d^2}\right) \prod_{i=1}^{n} \exp\!\left(-\frac{\lVert\vec z_{2i}' - \vec y_i\rVert^2 + \lVert\vec y_{2i+1}' - \vec y_i\rVert^2}{2w^2}\right),
\end{equation}
where $w$ is the width of the Gaussians from the previous stage.
For the first stage, a zeroth stage with just one Gaussian placed at the origin was assumed as the previous one. 
The step size $d$ was determined dynamically by slightly increasing or decreasing it after accepted or rejected steps, respectively. 

In each MCMC-step, also a new common width $\sigma$ of the Gaussians was proposed, with normally distributed proposals restricted to positive values. 
For the coliphage, also the heights $h_i$ of the Gaussians were determined.
Here, separate MCMC-steps with Gaussian proposals restricted to the uniform prior $\mathcal U(0.1, 1)$ were performed for the heights alternatingly with those for the positions, with an independently determined step size. 

\paragraph{Regularized likelihood function.}
For stages of reduced resolution, a regularized version of the likelihood function was used. 
To that end, consider a smoothed version of the true electron density function $\rho$ obtained by a convolution with a Gaussian kernel, $\tilde\rho = \rho*\mathcal N(\sigma_\mathrm{reg})$. 
The intensity function corresponding to this smoothed version is, due to the Fourier convolution theorem, given by the pointwise product of the original intensity function and the squared absolute value of the Fourier transform of the smoothing kernel, $I_{\tilde\rho}(\vec k) = I_\rho(\vec k) \cdot \exp(-\sigma_\mathrm{reg}^2\vec k^2)$.
This relationship was used to obtain the images that would have been generated for the smoothed structure by rejection sampling, which where then used in the likelihood instead of the original images for the stages of reduced resolution.

In the noise-free case, computational efficiency was further increased substantially by selecting only those original images for the likelihood computations that actually contain useful information at the respective resolution. As described in the Supplementary Text, the Bayesian formalism allows for removing this selection bias.

\paragraph{Structure alignment and resolution estimate.}
The resolution of the obtained electron densities was calculated using Fourier shell correlations \cite{van_heel_fourier_2005},
\begin{equation}
    \mathrm{FSC}(k, \rho_1, \rho_2) = \frac{\int_{\lVert\vec k\rVert = k} \hat\rho_1(\vec k)^*\hat\rho_2(\vec k)\,\mathrm d\vec k}{\sqrt{\int_{\lVert\vec k\rVert = k} |\hat\rho_1(\vec k)|^2\,\mathrm d\vec k}\sqrt{\int_{\lVert\vec k\rVert = k} |\hat\rho_2(\vec k)|^2\,\mathrm d\vec k}}\,,
\end{equation}
where $\rho_1$ and $\rho_2$ are the densities to be compared and $\hat\rho$ denotes the Fourier transform of $\rho$. The achieved resolution was determined as $2\pi/k_\mathrm{fsc}(\rho_1, \rho_2)$, where $k_\mathrm{fsc}(\rho_1, \rho_2)$ is the threshold at which the Fourier shell correlation drops below the conservative threshold of $0.5$ \cite{van_heel_fourier_2005}.
Because the orientations of the electron densities are random and irrelevant, they were aligned to each other before calculating the resolution, by maximizing $k_\mathrm{fsc}(\rho_1, \vec S\rho_2)$ over all orthogonal matrices $\vec S \in \mathrm{O}(3)$. Both rotations and reflections were included, as X-ray scattering images do not distinguish between mirror images. 
Here, $\vec S\rho$ denotes the rotated electron density obtained by applying $\vec S$ to all positions $\vec y_i$ from equation~\eqref{eq: representation}.

\clearpage

\printbibliography

\clearpage

\section*{Acknowledgments}
This work was financially supported by the Federal Ministry of Education and Research through the joint research project 05K20EGA Fluctuation XFEL, and the Deutsche Forschungsgemeinschaft (DFG, German Research Foundation) - CRC 1456/1 - 432680300. 

S.S. and H.G. conceived research, S.S. carried out research, S.S. and H.G. wrote paper. 

The authors declare that they have no competing interests.

All data needed to evaluate the conclusions in the paper are present in the paper and/or the Supplementary Materials.

We have implemented our method in the Julia programming language \cite{bezanson_julia_2017}. The source code is available at \href{https://gitlab.gwdg.de/sschult/xfel}{\url{https://gitlab.gwdg.de/sschult/xfel}}.


\clearpage

\renewcommand{\thetable}{S\arabic{table}}
\captionsetup{justification=raggedright,singlelinecheck=false,labelsep=newline,font=normalsize}
\setcounter{page}{1}

\begin{center}
    {\Large Supplementary Materials for: }\\[2ex]
    \textbf{\large Bayesian electron density determination from sparse and noisy single-molecule X-ray scattering images}\\[2ex]

    Steffen Schultze, Helmut Grubmüller$^*$

    {\small $^*$Corresponding author, Email: hgrubmu@mpinat.mpg.de}
\end{center}
\vspace{2cm}
\paragraph{This PDF file includes:}
\begin{itemize}
    \item Supplementary Text
    \item Table S1
\end{itemize}
\paragraph{Other Supplementary Materials for this manuscript include the following:}
\begin{itemize}
    \item Movie S1
\end{itemize}
\clearpage

\section*{Supplementary Text}
\subsection*{Computation}
\label{sup: computation}

We here describe the implementation of the likelihood computation for case of the noise-free model. 
The computations including noise and experimental effects are analogous, replacing, for example, the density of the Poisson distribution with that of the Gamma distribution. The source code is available at \href{https://gitlab.gwdg.de/sschult/xfel}{\url{https://gitlab.gwdg.de/sschult/xfel}}

The integral over $\mathrm{SO}(3)$ is approximated by a finite sum over rotations $\vec R_i$ with weights $w_i$,
\begin{equation}
    P(\vec k_1, \dots, \vec k_n \,|\, \rho) \approx \frac{N^n}{n!}\sum_i w_i\exp\left(-N\int_D I(\vec R_i\vec k) \,\mathrm d \vec k\right) \prod_{j=1}^n I(\vec R_i\vec k_j)
\end{equation}
Computing this sum involves evaluating the intensity function $I$ at all points of the form $\vec R_i\vec k_j$.
Since this has to be done for all the images, this leads to a very large number of evaluations of $I$. 
It is therefore efficient to first discretize the images. 
To that end, the detector is pixelated, that is, partitioned into a grid of cells with centers $\vec x_k$ and areas $a_k$. Each image $\vec k_1, \dots, \vec k_n$ is replaced with a set of indices $k_1, \dots, k_n$, such that for each $\vec k_i$ the closest point in the grid is $\vec x_{k_i}$. In this setting, the probability distribution becomes
\begin{equation}
    \label{eq:grids}
    P(k_1, \dots, k_n \,|\, \rho) \approx \frac{N^n}{n!}\sum_i w_i \exp\left(-N\sum_k a_k I(\vec R_i\vec x_k)\right) \prod_{j=1}^n a_{k_j} I(\vec R_i\vec x_{k_j})
\end{equation}

To construct the quadrature rule for $\mathrm{SO}(3)$, we proceed as follows. First, we choose a Lebedev grid as a uniform grid of points $\vec v_i$ in the 2-sphere $S^2$. For each one of these, we find a rotation $Q_i \in \mathrm{SO}(3)$ such that $Q_i\vec v_i \parallel \vec k_0$. In addition, let $S_j$ be uniformly spaced rotations around the axis defined by $\vec k_0$. The set of products $S_jQ_i$ is then a uniform grid in $\mathrm{SO}(3)$. 
The Lebedev precision and the number of angular rotations $S_j$ were chosen such that the expected angular distance between nearest neighbors in the resulting grid is smaller than the length scale corresponding to the desired relative resolution divided by the approximate radius of the molecule. Equation \eqref{eq:grids} becomes
\begin{equation}
    P(k_1, \dots, k_n \,|\, \rho) \approx \frac{N^n}{n!}\sum_{i,j} w_i\exp\left(-N \sum_k a_k I(Q_i S_j\vec x_k)\right) \prod_{m=1}^n a_{k_m} I(Q_iS_j\vec x_{k_m})
\end{equation}
Choosing the pixel grid $\vec x_k$ such that it is rotationally symmetric allows further simplification. We reindex it as $\vec x_{k,l}$, such that $S_j\vec x_{k,l} = \vec x_{k+j,l}$. Here, the first index is considered cyclic, that is, if, say, $k$ ranges from $1$ to $k_\mathrm{max}$, then $\vec x_{k+j,l}$ is to be interpreted as $\vec x_{(k+j\mod k_\mathrm{max}),l}$. The corresponding areas $a_{k,l}$ only depend on $l$, so we write $a_l = a_{k,l}$. The images now also consist of these new indices. Plugging this in, we get
\begin{align}
    P(k_1, l_1, \dots, k_n, l_n \,|\, \rho) &\approx \frac{N^n}{n!}\sum_{i,j} w_i\exp\left( -N \sum_{k,l} a_l I(Q_iS_j\vec x_{k,l})\right) \prod_{m=1}^n a_{l_m} I(Q_iS_j\vec x_{k_m,l_m})\\
    &= \frac{N^n}{n!}\sum_i w_i\exp\left( -N \sum_{k,l} a_l I(Q_i\vec x_{k,l})\right) \sum_j\prod_{m=1}^n a_{l_m} I(Q_i\vec x_{k_m+j,l_m}) \\
    &= \frac{N^n}{n!}\sum_i w_i P_i \sum_j\prod_{m=1}^n I_{i,k_m+j,l_m}
\end{align}
The values $I_{i,k,l} \coloneqq a_lI(Q_i\vec x_{k,l})$ and $P_i \coloneqq \exp(-N\sum_{k,l} I_{i,k,l})$ can be computed in advance and reused for each image.

Due to limited floating point precision, a number of adjustments must be made. Due to the large value of $N$, computing $P_i$ results in underflow. Therefore, we write \begin{equation}
    \tilde P_i = P_i /\bar P, \qquad \bar P =\left(\prod_{i'=1}^{i_\mathrm{max}} P_{i'}\right)^{\frac1{i_\mathrm{max}}}.
\end{equation}
Further, $I_{i,k,l} \ll 1$, so if the images contain enough photons the product over $m$ will underflow. Since the magnitude of $I_{i,k,l}$ depends mostly on $l$, we define
\begin{equation}
    \tilde I_{i,k,l} = I_{i,k,l} /\bar I_l, \qquad \bar I_l = \frac1{i_\mathrm{max}k_\mathrm{max}}\sum_{i'=1}^{i_\mathrm{max}}\sum_{k'=1}^{k_\mathrm{max}} I_{i',k',l}
\end{equation}
Both $\bar P$ and $\bar I_l$ do not depend on the rotation index $i$ and factor out,
\begin{equation}
    P(k_1, l_1, \dots, k_n, l_n \,|\, \rho) \approx \frac{N^n}{n!}\bar P\left(\prod_{m=1}^n\bar I_{l_m}\right)\sum_i w_i \tilde P_i \sum_j\prod_{m=1}^n \tilde I_{i,k_m+j,l_m}
\end{equation}
Taking the logarithm,
\begin{equation}
    \log P(k_1, l_1, \dots, k_n, l_n \,|\, \rho) \approx \log \frac{N^n}{n!} + \log\bar P + \sum_{m=1}^n\log \bar I_{l_m}
    +\log \sum_i w_i \tilde P_i \sum_j\prod_{m=1}^n \tilde I_{i,k_m+j,l_m},
\end{equation}
we see that only $\log \bar P$ and $\log \bar I_l$ appear, which can be computed without overflow.

\subsection*{Image selection}
\label{sup: selection}
In our hierarchical sampling scheme, images containing only photons with $k$ below a threshold are no longer useful, and the computations were sped up by removing these images. 
To achieve this, only the subset $\mathcal I_C$ of those images was used that fulfilled the condition $C(I)$ that for each $i$ the image $I$ contains at least $m_i$ photons with $r_i < k < r_{i+1}$, where the parameters $r_i$ and $m_i$ were chosen such that the radial distribution of photons in the selected images was close to uniform up to the desired resolution level (Table \ref{tab: parameters}).
To ensure that the posterior was not biased by this filtering, it was taken into account in the Bayesian formalism by dividing by the probability $P(C\,|\,\rho)$ that an image fulfills $C$. In other words, the original posterior probability was replaced with $P(\rho\,|\,\mathcal I_C, C) \propto P(\mathcal I_C \,|\, \rho) / P(C\,|\,\rho)$. The probability that an image fulfills $C$ depends on the orientation $\vec R$. Therefore, $P(C\,|\,\rho)$ was obtained by averaging,
\begin{equation}
    P(C\,|\,\rho) = \int_{\mathrm{SO}(3)} \prod_i \left(1 - Q\left(m_i-1, N\!\!\int_{D_i}\!\!\mathcal \lvert F\{\rho\}(\vec R\vec k)\rvert^2 \,\mathrm d \vec k \right)\right) \mathrm d \vec R,
\end{equation}
where $Q(x, \lambda)$ is the cumulative distribution function of a Poisson distribution with mean $\lambda$ and $D_i = \{\vec k \in D \,|\, r_i < \lVert \vec k \rVert < r_{i+1}\}$ is the relevant slice of the Ewald sphere.

\clearpage

\addtolength{\tabcolsep}{-2.1pt}

\begin{table}[ht!]
    \footnotesize
    \centering
    \pgfplotstabletypeset[
        col sep=&, row sep=\\,
        every head row/.style={before row = \toprule},
        every last row/.style={after row = \bottomrule},
    	columns/name/.style={column name = {Test Case}, string type},
    	columns/stage/.style={column name = {Stage}},
    	columns/total/.style={column name = {\shortstack{total\\images}}, sci, sci zerofill, precision=2},
    	columns/selected/.style={column name = {\shortstack{selected\\images}}, fixed, empty cells with=-},
    	columns/nmask/.style={column name = {$n_i$}, string type, empty cells with=-},
    	columns/rmask1/.style={column name = {$r_i\,\,[1/\text{{\AA}}$]}, string type, empty cells with=-},
    	columns/rmask2/.style={column name = {$b_i$}, string type},
    	columns/kwidth/.style={column name = {$\sigma_\mathrm{reg}\,\,[\text{{\AA}}]$}, fixed, fixed zerofill, precision = 1},
    	columns/T0/.style={column name = {$T_0$}},
    	columns/thalf/.style={column name = {$t_{1/2}$}, sci, precision=1},
    	columns/ngauss/.style={column name = {\shortstack{number of \\ Gaussians}}},
       	columns/precision/.style={column name = {{\shortstack{Lebedev\\precision}}}},
    	columns/npercircle/.style={column name = {{\shortstack{angular\\rotations}}}},
    ]{
        name & stage & total & selected & nmask & rmask1 & kwidth & thalf & ngauss & precision & npercircle \\
        \midrule
        \multirow{5}{*}{\shortstack{Crambin \\ noise-free}} & 1 & 8956 & 1000 & 4 & $0.25, \infty$ & 2.0 & 1000 & 12 & 23 & 32 \\
        & 2 & 1e7 & 19315 & 3, 2 & $0.33, 0.5, \infty$ & 1.5 & 10000 & 23 & 47 & 32 \\
        & 3 & 3044813 & 50000 & 1, 1, 2 & $0.35, 0.5, 0.65, \infty$ & 1.2 & 20000 & 46 & 47 & 64 \\
        & 4 & 1e8 & 204447 & 1, 2, 2 & $0.35, 0.5, 0.8, \infty$ & 0.9 & 100000 & 92 & 89 & 64 \\
        & 5 & 1e8 & 634032 & 1, 1, 3 & $0.4, 0.65, 0.9, \infty$ & 0.5 & 100000 & 184 & 89 & 64 \\
        \midrule
        \multirow{2}{*}{\shortstack{Crambin \\ $75\,\%$ noise}} & 1 & 1000000 &  &  &  & 1 & 500 & 12 & 23 & 32 \\
        & 2 & 1000000 &  &  &  & 0.0 & 1e4 & 23 & 47 & 64 \\
        \midrule
        \shortstack{Crambin \\ $90\,\%$ noise} & 1 & 3000000 &  &  &  & 1 & 500 & 12 & 23 & 32 \\
        \midrule
        \multirow{2}{*}{\shortstack{Coliphage\\PR772}} & 1 & 1500 &  &  &  & 0 & 10000 & 13 & 23 & 64 \\
        & 2 & 200000 &  &  &  & 0 & 50000 & 200 & 23 & 64 \\
    }
    \caption{Parameters used for each of the test cases.}
    \label{tab: parameters}
\end{table}

\clearpage
\textbf{Movie S1.}\\
Rotating 3D isosurface of the averaged determined electron density for coliphage PR772. Perfect icosahedron of side length $30\,\mathrm{nm}$ for reference. Isosurface drawn at $5\,\%$ of the maximum value of $\rho$.

\end{document}